\documentclass[amsmath,amssymb,amsbsy,prl,showpacs,floatfix,twocolumn]{revtex4-1}

\usepackage{amsmath}
\usepackage{amssymb}
\usepackage{bm}
\usepackage{xcolor}
\usepackage[utf8]{inputenc}
\usepackage{graphicx}
\usepackage{dsfont} 
\usepackage{subfigure}
\usepackage{wasysym}
\usepackage{hyperref}
\usepackage{setspace}

\begin{document}

\title{Topology in abundance 
}
\author{\'{E}tienne Lantagne-Hurtubise}
\author{Marcel Franz}
\affiliation{Department of Physics and Astronomy \& Quantum Matter Institute, University of British Columbia, Vancouver, BC, Canada V6T 1Z1}

\begin{abstract}
The topological classification of all known non-magnetic crystalline compounds is now complete, revealing thousands of new candidate topological materials waiting to be explored in the lab.
\end{abstract}

\date{\today}

\maketitle

Topological materials show exotic phenomenology, such as protected surface states and quantized electromagnetic and thermal responses, which are directly related to \emph{topological invariants} -- quantities that only depend on the global properties of the material’s band structure. Yet the traditional way of calculating such topological invariants is challenging, thus hindering its widespread application to a large number of materials. Building on recent theoretical breakthroughs which provide a systematic mapping between the crystal symmetries of a material and its topological properties, three independent research groups \cite{Vergniory2019,Tang2019,Zhang2019} have completed a catalogue of all ($>26,000$) nonmagnetic crystalline compounds available in the International Crystal Structure Database (ICSD). Writing in Nature, they predict a treasure trove containing thousands of new topological materials which should keep experimentalists busy for many years to come. User-friendly, open-access online tools allow for quick verification of the topological properties of any given compound.

In the past 14 years, tremendous effort has been devoted to the study of topological phases of matter. A hallmark result is the classification of all gapped, non-interacting electronic phases with internal symmetries (time-reversal, particle-hole and chirality) as contained in the periodic table of topological insulators and superconductors \cite{Kitaev2009}. The subsequent addition of spatial symmetries has expanded these possibilities with the class of topological crystalline insulators, which encompass the recently discovered higher-order topological insulators. The overarching goal of this line of work is to enumerate all the \emph{possible} topological phases, given the symmetries and dimensionality of the system. However, such classifications do little to inform us on \emph{how} to obtain topological phases -- i.e., which model or material might realize them. This situation has changed in the past two years with two independent developments showing how the topological properties of a particular compound can be inferred using only its symmetries and orbital content. 
\begin{figure*}
\centering
\includegraphics[width=0.7\textwidth]{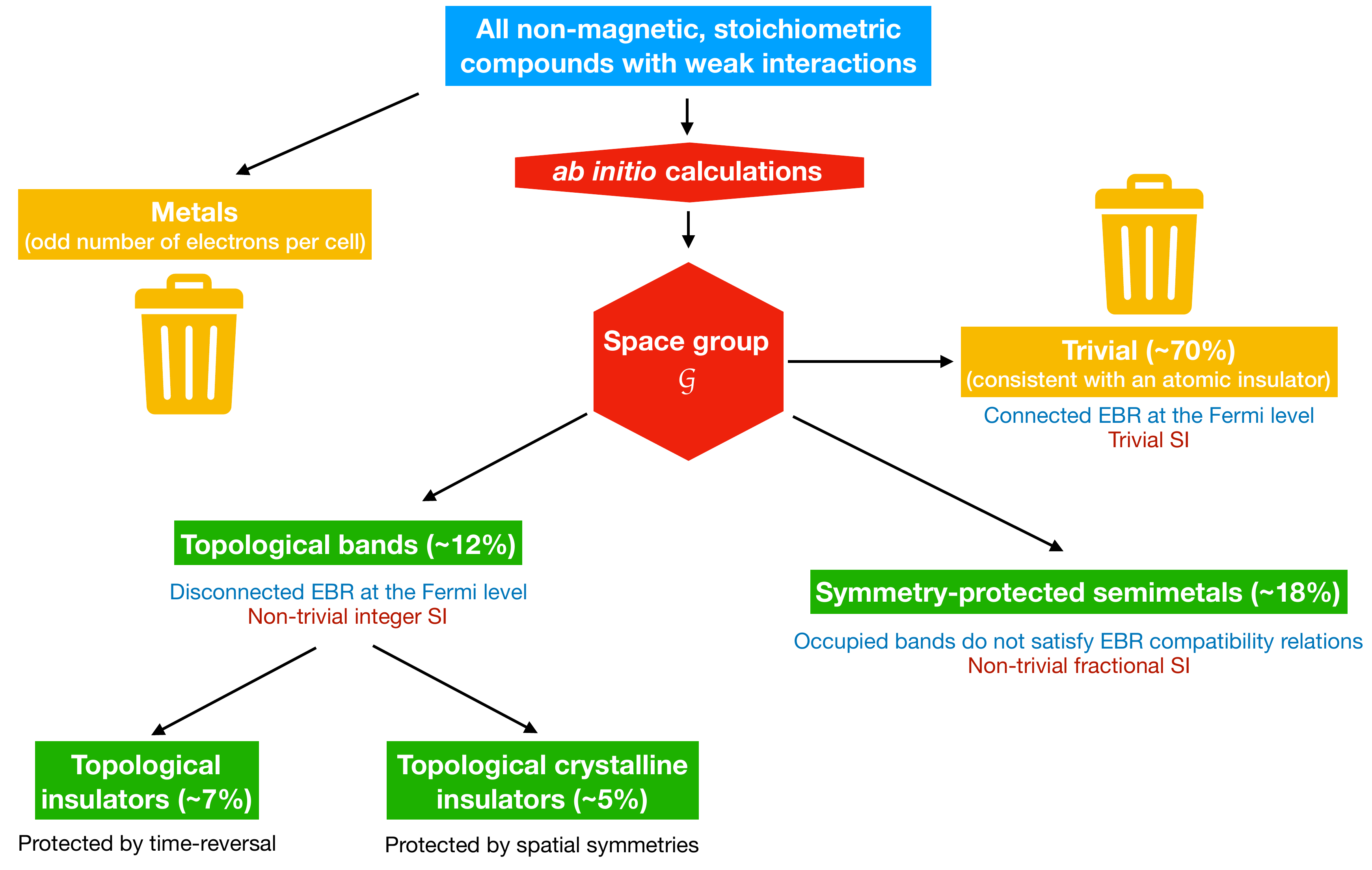}
\caption{Diagram representing the materials search in Refs.~\cite{Vergniory2019,Tang2019,Zhang2019} which classifies \emph{ab initio} band structures into three main categories: ``topological bands", ``symmetry-protected semimetals" and ``consistent with an atomic insulator". The diagnostics criteria for each category are shown in blue for the band-representation technique, and in red for the symmetry-indicator technique. EBR stands for elementary band representation and SI denotes symmetry indicators. The percentages in each category are taken from Refs.~\cite{Vergniory2019, Zhang2019}. Note that Ref.~\cite{Tang2019} reports a much larger proportion of topological compounds ($\sim 57\%$), presumably because their analysis also includes bands far from the Fermi level.\label{fig1}}
\end{figure*}

The first approach, pioneered by Bradlyn et al.~\cite{Bradlyn2017}, builds on the notion of band representation introduced almost forty years ago by Zak \cite{Zak1980}. A band representation encodes the symmetry properties of a band (or group of bands) arising from a set of localized orbitals respecting the crystal symmetries, and possibly time-reversal. A band representation thus describes bands that are consistent with a trivial atomic insulator. Topological bands are then defined as bands that are \emph{not} described by a band representation, i.e. they do not admit a localized atomic limit which respects the crystal symmetries. In more technical terms, topological bands are described by a \emph{disconnected} elementary band representation (EBR) \cite{Bradlyn2017, Cano2018a}. Interestingly, such a possibility was originally ruled out by Michel and Zak \cite{Michel2001}, who only considered single-valued representations occuring when the electron spin is neglected. Generalizing the theory of band representations to spinful fermions and accounting for spin-orbit coupling was a crucial step in establishing the results of Refs.~\cite{Bradlyn2017, Cano2018a} and enabling the materials search of Ref.~\cite{Vergniory2019}. 

The second approach is based on the concept of symmetry indicators (SI) \cite{Po2017}. This is inspired by the early work of Fu and Kane \cite{Fu2007} who realized that, for topological insulators (TIs) with inversion symmetry, the Z$_2$ topological invariant could be obtained simply from the eigenvalues of the inversion operator at a few high-symmetry points in the Brillouin zone. Symmetry indicators are a generalization of this idea -- classifying topology from symmetry eigenvalues -- to an arbitrary space group $\mathcal{G}$. In essence, they diagnose how the symmetry content of the wavefunctions at high-symmetry points \emph{differs} from that of a trivial atomic insulator. This method was developed by Po et al. \cite{Po2017} for all 230 space groups in 3D with spin-orbit coupling, partly based on pioneering ideas by Kruthoff et al. \cite{Kruthoff}. Shortly thereafter, the full mapping between symmetry indicators and the standard topological invariants (which in general is not one-to-one) was reported~\cite{Song2018, Khalaf2018}. Symmetry indicators form the core of the materials search of Refs.~\cite{Tang2019,Zhang2019}.

Applying either theory to concrete materials first requires performing a routine \emph{ab initio} band structure calculation, out of which the symmetry properties of Bloch wavefunctions at high-symmetry points are extracted. This data is then fed into a series of algorithms which classify the band structure into three main categories: (i) consistent with an atomic insulator, (ii) symmetry-protected semimetal, or (iii) topological bands. A cartoon of this procedure is shown in Fig.~\ref{fig1}. Many further subdivisions of the three main categories are possible and differ slightly between the different papers~\cite{Vergniory2019, Tang2019, Zhang2019}.

The most surprising conclusion of this search is the sheer abundance of topological materials in nature. It is now estimated that  $27 - 30\%$ of all non-magnetic crystalline materials in ICSD are topological, with roughly $12\%$ insulators and $15 - 18\%$ semimetals (see Fig.~\ref{fig1}). This represents around 8000 new topological compounds! For nearly all the materials identified, the three papers report a wealth of experimentally useful information. For topological (crystalline) insulators, this includes the protecting symmetries, the size of the topological gap and the potential presence of trivial bands within it that might mask topological properties. For topological semimetals, this includes whether the protected crossing occurs at a high-symmetry point or line, the potential presence of nodal rings, or even ways to open a topological gap at the Fermi level by applying symmetry-breaking perturbations \cite{Vergniory2019}. 

A mildly disappointing aspect of these results is that no obvious “ideal” topological material -- or at least one clearly better than the currently known examples --  has emerged. For example, out of  $\sim3000$ new TI candidates, it appears that Bi$_2$Se$_3$ -- one of the first TIs discovered -- still has the largest bandgap. Further work will be necessary to fully explore the list of candidates in hope of finding a hidden gem. 

For the experimental community, the most useful outcome of this work is perhaps the construction of two online tools allowing to quickly ascertain the topological properties of a compound of interest. In this regard, Ref.~\cite{Zhang2019} introduced a user-friendly \href{http://materiae.iphy.ac.cn/}{interactive tool} in the form of a periodic table. Ref.~\cite{Vergniory2019} incorporated a routine to the \href{http://www.cryst.ehu.es/cgi-bin/cryst/programs/topological.pl}{Bilbao Crystallographic Server} which takes as an input the result of an \emph{ab initio} band structure calculation and outputs the topological characteristics of the material. These tools promise to considerably speed up and democratize the process of predicting and verifying topological materials.

A few words of caution are in order when interpreting the gigantic databases reported in Refs.~\cite{Zhang2019, Tang2019, Vergniory2019}. First, it is important to note that both techniques rely on symmetry representations along high-symmetry points and lines in the Brillouin zone. As such, they are insensitive to any \emph{accidental} band crossings (as occur in Weyl semimetals) or band inversions at generic momenta. Second, both approaches need input from \emph{ab initio} density functional theory calculations to determine the system energetics (i.e.\ the energy ordering of the bands). This can be problematic for small bandgap materials where density functional theory might get the ordering of the bands wrong. For the same reason, reported gap values should be treated with a healthy dose of skepticism. Third, the classification scheme is based on the independent electron approximation and will fail when correlation effects become large. This might be important especially when considering the $3d$, $4f$ and $5f$ compounds included in the databases. Finally, the number of materials reported in each category differs slightly between the different approaches and papers. Reconciling these numbers by thorough cross checks will be crucial to get a clear picture of the topological materials landscape. 

An exciting new direction will be to generalize this search to magnetic materials. This task is complicated by the large number of magnetic space groups (1651) and their different group-theoretic properties, but important steps in this direction have already been taken \cite{Watanabe2018}. A sweep of magnetic materials in ICSD has not yet been reported, but should be completed in the near future and reveal many more topological compounds including, perhaps, a few surprises.

\bibliography{biblio}

\begin{thebibliography}{14}%
\makeatletter
\providecommand \@ifxundefined [1]{%
 \@ifx{#1\undefined}
}%
\providecommand \@ifnum [1]{%
 \ifnum #1\expandafter \@firstoftwo
 \else \expandafter \@secondoftwo
 \fi
}%
\providecommand \@ifx [1]{%
 \ifx #1\expandafter \@firstoftwo
 \else \expandafter \@secondoftwo
 \fi
}%
\providecommand \natexlab [1]{#1}%
\providecommand \enquote  [1]{``#1''}%
\providecommand \bibnamefont  [1]{#1}%
\providecommand \bibfnamefont [1]{#1}%
\providecommand \citenamefont [1]{#1}%
\providecommand \href@noop [0]{\@secondoftwo}%
\providecommand \href [0]{\begingroup \@sanitize@url \@href}%
\providecommand \@href[1]{\@@startlink{#1}\@@href}%
\providecommand \@@href[1]{\endgroup#1\@@endlink}%
\providecommand \@sanitize@url [0]{\catcode `\\12\catcode `\$12\catcode
  `\&12\catcode `\#12\catcode `\^12\catcode `\_12\catcode `\%12\relax}%
\providecommand \@@startlink[1]{}%
\providecommand \@@endlink[0]{}%
\providecommand \url  [0]{\begingroup\@sanitize@url \@url }%
\providecommand \@url [1]{\endgroup\@href {#1}{\urlprefix }}%
\providecommand \urlprefix  [0]{URL }%
\providecommand \Eprint [0]{\href }%
\providecommand \doibase [0]{http://dx.doi.org/}%
\providecommand \selectlanguage [0]{\@gobble}%
\providecommand \bibinfo  [0]{\@secondoftwo}%
\providecommand \bibfield  [0]{\@secondoftwo}%
\providecommand \translation [1]{[#1]}%
\providecommand \BibitemOpen [0]{}%
\providecommand \bibitemStop [0]{}%
\providecommand \bibitemNoStop [0]{.\EOS\space}%
\providecommand \EOS [0]{\spacefactor3000\relax}%
\providecommand \BibitemShut  [1]{\csname bibitem#1\endcsname}%
\let\auto@bib@innerbib\@empty
\bibitem [{\citenamefont {Vergniory}\ \emph {et~al.}(2019)\citenamefont
  {Vergniory}, \citenamefont {Elcoro}, \citenamefont {Felser}, \citenamefont
  {Regnault}, \citenamefont {Bernevig},\ and\ \citenamefont
  {Wang}}]{Vergniory2019}%
  \BibitemOpen
  \bibfield  {author} {\bibinfo {author} {\bibfnamefont {M.~G.}\ \bibnamefont
  {Vergniory}}, \bibinfo {author} {\bibfnamefont {L.}~\bibnamefont {Elcoro}},
  \bibinfo {author} {\bibfnamefont {C.}~\bibnamefont {Felser}}, \bibinfo
  {author} {\bibfnamefont {N.}~\bibnamefont {Regnault}}, \bibinfo {author}
  {\bibfnamefont {B.~A.}\ \bibnamefont {Bernevig}}, \ and\ \bibinfo {author}
  {\bibfnamefont {Z.}~\bibnamefont {Wang}},\ }\href {\doibase
  10.1038/s41586-019-0954-4} {\bibfield  {journal} {\bibinfo  {journal}
  {Nature}\ }\textbf {\bibinfo {volume} {566}},\ \bibinfo {pages} {480}
  (\bibinfo {year} {2019})}\BibitemShut {NoStop}%
\bibitem [{\citenamefont {Tang}\ \emph {et~al.}(2019)\citenamefont {Tang},
  \citenamefont {Po}, \citenamefont {Vishwanath},\ and\ \citenamefont
  {Wan}}]{Tang2019}%
  \BibitemOpen
  \bibfield  {author} {\bibinfo {author} {\bibfnamefont {F.}~\bibnamefont
  {Tang}}, \bibinfo {author} {\bibfnamefont {H.~C.}\ \bibnamefont {Po}},
  \bibinfo {author} {\bibfnamefont {A.}~\bibnamefont {Vishwanath}}, \ and\
  \bibinfo {author} {\bibfnamefont {X.}~\bibnamefont {Wan}},\ }\href {\doibase
  10.1038/s41586-019-0937-5} {\bibfield  {journal} {\bibinfo  {journal}
  {Nature}\ }\textbf {\bibinfo {volume} {566}},\ \bibinfo {pages} {486}
  (\bibinfo {year} {2019})}\BibitemShut {NoStop}%
\bibitem [{\citenamefont {Zhang}\ \emph {et~al.}(2019)\citenamefont {Zhang},
  \citenamefont {Jiang}, \citenamefont {Song}, \citenamefont {Huang},
  \citenamefont {He}, \citenamefont {Fang}, \citenamefont {Weng},\ and\
  \citenamefont {Fang}}]{Zhang2019}%
  \BibitemOpen
  \bibfield  {author} {\bibinfo {author} {\bibfnamefont {T.}~\bibnamefont
  {Zhang}}, \bibinfo {author} {\bibfnamefont {Y.}~\bibnamefont {Jiang}},
  \bibinfo {author} {\bibfnamefont {Z.}~\bibnamefont {Song}}, \bibinfo {author}
  {\bibfnamefont {H.}~\bibnamefont {Huang}}, \bibinfo {author} {\bibfnamefont
  {Y.}~\bibnamefont {He}}, \bibinfo {author} {\bibfnamefont {Z.}~\bibnamefont
  {Fang}}, \bibinfo {author} {\bibfnamefont {H.}~\bibnamefont {Weng}}, \ and\
  \bibinfo {author} {\bibfnamefont {C.}~\bibnamefont {Fang}},\ }\href {\doibase
  10.1038/s41586-019-0944-6} {\bibfield  {journal} {\bibinfo  {journal}
  {Nature}\ }\textbf {\bibinfo {volume} {566}},\ \bibinfo {pages} {475}
  (\bibinfo {year} {2019})}\BibitemShut {NoStop}%
\bibitem [{\citenamefont {Kitaev}(2009)}]{Kitaev2009}%
  \BibitemOpen
  \bibfield  {author} {\bibinfo {author} {\bibfnamefont {A.}~\bibnamefont
  {Kitaev}},\ }\href {\doibase 10.1063/1.3149495} {\bibfield  {journal}
  {\bibinfo  {journal} {AIP Conference Proceedings}\ }\textbf {\bibinfo
  {volume} {1134}},\ \bibinfo {pages} {22} (\bibinfo {year}
  {2009})}\BibitemShut {NoStop}%
\bibitem [{\citenamefont {Bradlyn}\ \emph {et~al.}(2017)\citenamefont
  {Bradlyn}, \citenamefont {Elcoro}, \citenamefont {Cano}, \citenamefont
  {Vergniory}, \citenamefont {Wang}, \citenamefont {Felser}, \citenamefont
  {Aroyo},\ and\ \citenamefont {Bernevig}}]{Bradlyn2017}%
  \BibitemOpen
  \bibfield  {author} {\bibinfo {author} {\bibfnamefont {B.}~\bibnamefont
  {Bradlyn}}, \bibinfo {author} {\bibfnamefont {L.}~\bibnamefont {Elcoro}},
  \bibinfo {author} {\bibfnamefont {J.}~\bibnamefont {Cano}}, \bibinfo {author}
  {\bibfnamefont {M.~G.}\ \bibnamefont {Vergniory}}, \bibinfo {author}
  {\bibfnamefont {Z.}~\bibnamefont {Wang}}, \bibinfo {author} {\bibfnamefont
  {C.}~\bibnamefont {Felser}}, \bibinfo {author} {\bibfnamefont {M.~I.}\
  \bibnamefont {Aroyo}}, \ and\ \bibinfo {author} {\bibfnamefont {B.~A.}\
  \bibnamefont {Bernevig}},\ }\href {\doibase 10.1038/nature23268} {\bibfield
  {journal} {\bibinfo  {journal} {Nature}\ }\textbf {\bibinfo {volume} {547}},\
  \bibinfo {pages} {298} (\bibinfo {year} {2017})}\BibitemShut {NoStop}%
\bibitem [{\citenamefont {Zak}(1980)}]{Zak1980}%
  \BibitemOpen
  \bibfield  {author} {\bibinfo {author} {\bibfnamefont {J.}~\bibnamefont
  {Zak}},\ }\href {\doibase 10.1103/PhysRevLett.45.1025} {\bibfield  {journal}
  {\bibinfo  {journal} {Phys. Rev. Lett.}\ }\textbf {\bibinfo {volume} {45}},\
  \bibinfo {pages} {1025} (\bibinfo {year} {1980})}\BibitemShut {NoStop}%
\bibitem [{\citenamefont {Cano}\ \emph {et~al.}(2018)\citenamefont {Cano},
  \citenamefont {Bradlyn}, \citenamefont {Wang}, \citenamefont {Elcoro},
  \citenamefont {Vergniory}, \citenamefont {Felser}, \citenamefont {Aroyo},\
  and\ \citenamefont {Bernevig}}]{Cano2018a}%
  \BibitemOpen
  \bibfield  {author} {\bibinfo {author} {\bibfnamefont {J.}~\bibnamefont
  {Cano}}, \bibinfo {author} {\bibfnamefont {B.}~\bibnamefont {Bradlyn}},
  \bibinfo {author} {\bibfnamefont {Z.}~\bibnamefont {Wang}}, \bibinfo {author}
  {\bibfnamefont {L.}~\bibnamefont {Elcoro}}, \bibinfo {author} {\bibfnamefont
  {M.~G.}\ \bibnamefont {Vergniory}}, \bibinfo {author} {\bibfnamefont
  {C.}~\bibnamefont {Felser}}, \bibinfo {author} {\bibfnamefont {M.~I.}\
  \bibnamefont {Aroyo}}, \ and\ \bibinfo {author} {\bibfnamefont {B.~A.}\
  \bibnamefont {Bernevig}},\ }\href {\doibase 10.1103/PhysRevB.97.035139}
  {\bibfield  {journal} {\bibinfo  {journal} {Phys. Rev. B}\ }\textbf {\bibinfo
  {volume} {97}},\ \bibinfo {pages} {035139} (\bibinfo {year}
  {2018})}\BibitemShut {NoStop}%
\bibitem [{\citenamefont {Michel}\ and\ \citenamefont
  {Zak}(2001)}]{Michel2001}%
  \BibitemOpen
  \bibfield  {author} {\bibinfo {author} {\bibfnamefont {L.}~\bibnamefont
  {Michel}}\ and\ \bibinfo {author} {\bibfnamefont {J.}~\bibnamefont {Zak}},\
  }\href {\doibase 10.1016/S0370-1573(00)00093-4} {\bibfield  {journal}
  {\bibinfo  {journal} {Physics Reports}\ }\textbf {\bibinfo {volume} {341}},\
  \bibinfo {pages} {377 } (\bibinfo {year} {2001})}\BibitemShut {NoStop}%
\bibitem [{\citenamefont {Po}\ \emph {et~al.}(2017)\citenamefont {Po},
  \citenamefont {Vishwanath},\ and\ \citenamefont {Watanabe}}]{Po2017}%
  \BibitemOpen
  \bibfield  {author} {\bibinfo {author} {\bibfnamefont {H.~C.}\ \bibnamefont
  {Po}}, \bibinfo {author} {\bibfnamefont {A.}~\bibnamefont {Vishwanath}}, \
  and\ \bibinfo {author} {\bibfnamefont {H.}~\bibnamefont {Watanabe}},\ }\href
  {https://www.nature.com/articles/s41467-017-00133-2} {\bibfield  {journal}
  {\bibinfo  {journal} {Nature Communications}\ } (\bibinfo {year}
  {2017})}\BibitemShut {NoStop}%
\bibitem [{\citenamefont {Fu}\ and\ \citenamefont {Kane}(2007)}]{Fu2007}%
  \BibitemOpen
  \bibfield  {author} {\bibinfo {author} {\bibfnamefont {L.}~\bibnamefont
  {Fu}}\ and\ \bibinfo {author} {\bibfnamefont {C.~L.}\ \bibnamefont {Kane}},\
  }\href {\doibase 10.1103/PhysRevB.76.045302} {\bibfield  {journal} {\bibinfo
  {journal} {Phys. Rev. B}\ }\textbf {\bibinfo {volume} {76}},\ \bibinfo
  {pages} {045302} (\bibinfo {year} {2007})}\BibitemShut {NoStop}%
\bibitem [{\citenamefont {Kruthoff}\ \emph {et~al.}(2017)\citenamefont
  {Kruthoff}, \citenamefont {de~Boer}, \citenamefont {van Wezel}, \citenamefont
  {Kane},\ and\ \citenamefont {Slager}}]{Kruthoff}%
  \BibitemOpen
  \bibfield  {author} {\bibinfo {author} {\bibfnamefont {J.}~\bibnamefont
  {Kruthoff}}, \bibinfo {author} {\bibfnamefont {J.}~\bibnamefont {de~Boer}},
  \bibinfo {author} {\bibfnamefont {J.}~\bibnamefont {van Wezel}}, \bibinfo
  {author} {\bibfnamefont {C.~L.}\ \bibnamefont {Kane}}, \ and\ \bibinfo
  {author} {\bibfnamefont {R.-J.}\ \bibnamefont {Slager}},\ }\href {\doibase
  10.1103/PhysRevX.7.041069} {\bibfield  {journal} {\bibinfo  {journal} {Phys.
  Rev. X}\ }\textbf {\bibinfo {volume} {7}},\ \bibinfo {pages} {041069}
  (\bibinfo {year} {2017})}\BibitemShut {NoStop}%
\bibitem [{\citenamefont {Song}\ \emph {et~al.}(2018)\citenamefont {Song},
  \citenamefont {Zhang}, \citenamefont {Fang},\ and\ \citenamefont
  {Fang}}]{Song2018}%
  \BibitemOpen
  \bibfield  {author} {\bibinfo {author} {\bibfnamefont {Z.}~\bibnamefont
  {Song}}, \bibinfo {author} {\bibfnamefont {T.}~\bibnamefont {Zhang}},
  \bibinfo {author} {\bibfnamefont {Z.}~\bibnamefont {Fang}}, \ and\ \bibinfo
  {author} {\bibfnamefont {C.}~\bibnamefont {Fang}},\ }\href
  {https://doi.org/10.1038/s41467-018-06010-w} {\bibfield  {journal} {\bibinfo
  {journal} {Nature Communications}\ }\textbf {\bibinfo {volume} {9}} (\bibinfo
  {year} {2018})}\BibitemShut {NoStop}%
\bibitem [{\citenamefont {Khalaf}\ \emph {et~al.}(2018)\citenamefont {Khalaf},
  \citenamefont {Po}, \citenamefont {Vishwanath},\ and\ \citenamefont
  {Watanabe}}]{Khalaf2018}%
  \BibitemOpen
  \bibfield  {author} {\bibinfo {author} {\bibfnamefont {E.}~\bibnamefont
  {Khalaf}}, \bibinfo {author} {\bibfnamefont {H.~C.}\ \bibnamefont {Po}},
  \bibinfo {author} {\bibfnamefont {A.}~\bibnamefont {Vishwanath}}, \ and\
  \bibinfo {author} {\bibfnamefont {H.}~\bibnamefont {Watanabe}},\ }\href
  {\doibase 10.1103/PhysRevX.8.031070} {\bibfield  {journal} {\bibinfo
  {journal} {Phys. Rev. X}\ }\textbf {\bibinfo {volume} {8}},\ \bibinfo {pages}
  {031070} (\bibinfo {year} {2018})}\BibitemShut {NoStop}%
\bibitem [{\citenamefont {Watanabe}\ \emph {et~al.}(2018)\citenamefont
  {Watanabe}, \citenamefont {Po},\ and\ \citenamefont
  {Vishwanath}}]{Watanabe2018}%
  \BibitemOpen
  \bibfield  {author} {\bibinfo {author} {\bibfnamefont {H.}~\bibnamefont
  {Watanabe}}, \bibinfo {author} {\bibfnamefont {H.~C.}\ \bibnamefont {Po}}, \
  and\ \bibinfo {author} {\bibfnamefont {A.}~\bibnamefont {Vishwanath}},\
  }\href {\doibase 10.1126/sciadv.aat8685} {\bibfield  {journal} {\bibinfo
  {journal} {Science Advances}\ }\textbf {\bibinfo {volume} {4}},\ \bibinfo
  {pages} {eaat8685} (\bibinfo {year} {2018})}\BibitemShut {NoStop}%
\end{thebibliography}%

\end{document}